\documentclass[sigconf]{acmart}

\usepackage{breakurl}

\acmYear{2020}\copyrightyear{2020}
\setcopyright{rightsretained}
\acmConference[FAT* '20]{Conference on Fairness, Accountability, and Transparency}{January 27--30, 2020}{Barcelona, Spain}
\acmBooktitle{Conference on Fairness, Accountability, and Transparency (FAT* '20), January 27--30, 2020, Barcelona, Spain}
\acmPrice{}
\acmDOI{10.1145/3351095.3372871}
\acmISBN{978-1-4503-6936-7/20/01}


\begin{document}

\title{Roles for Computing in Social Change}

\author{Rediet Abebe}
\affiliation{%
  \institution{Harvard University}}
\email{rabebe@fas.harvard.edu}

\author{Solon Barocas}
\affiliation{%
  \institution{Microsoft Research and Cornell University}}
\email{sbarocas@cornell.edu}

\author{Jon Kleinberg}
\affiliation{%
  \institution{Cornell University}}
\email{kleinber@cs.cornell.edu}

\author{Karen Levy}
\affiliation{%
  \institution{Cornell University}}
\email{karen.levy@cornell.edu}

\author{Manish Raghavan}
\affiliation{%
  \institution{Cornell University}}
\email{manish@cs.cornell.edu}

\author{David G. Robinson}
\affiliation{%
  \institution{Cornell University}}
\email{david.robinson@cornell.edu}

\begin{abstract}

A recent normative turn in computer science has brought concerns about fairness, bias, and accountability to the core of the field. Yet recent scholarship has warned that much of this technical work treats problematic features of the status quo as fixed, and fails to address deeper patterns of injustice and inequality. While acknowledging these critiques, we posit that computational research has valuable roles to play in addressing social problems --- roles whose value can be recognized even from a perspective that aspires toward fundamental social change. In this paper, we articulate four such roles, through an analysis that considers the opportunities as well as the significant risks inherent in such work. Computing research can serve as a \textit{diagnostic}, helping us to understand and measure social problems with precision and clarity. As a \textit{formalizer}, computing shapes how social problems are explicitly defined --- changing how those problems, and possible responses to them, are understood. Computing serves as \textit{rebuttal} when it illuminates the boundaries of what is possible through technical means. And computing acts as \textit{synecdoche} when it makes long-standing social problems newly salient in the public eye. We offer these paths forward as modalities that leverage the particular strengths of computational work in the service of social change, without overclaiming computing's capacity to solve social problems on its own.

\end{abstract}


\maketitle

\section{Introduction}

In high-stakes decision-making, algorithmic systems have the potential to predict outcomes more accurately and to allocate scarce societal resources more efficiently --- but also the potential to introduce, perpetuate, and worsen inequality. Algorithmic accountability, fairness, and bias have quickly become watchwords of mainstream technical research and practice, gaining currency both in computer science scholarship and in technology companies. And these concerns are having a public moment, with policymakers and practitioners newly attuned to their import.

Recently, these concerns have sparked debate about the relationship between
computing and social change --- in particular, about the degree to which
technical interventions can address fundamental problems of justice and equity.
Scholars have recently raised concerns about taking a computational lens to
certain social problems, questioning whether modifications to automated
decision-making can ever address the structural conditions that relegate certain
social groups to the margins of society~\cite{eubanks2018,crawford2018,gebru2019,broussard2019,green2019data,dignazio2019}. On these accounts, realizing principles of justice and equity requires addressing the root social, economic, and political origins of these problems --- not optimizing around them~\cite{barabas2017,costanza2018,whittaker2018,pasquale2018,powles2018,selbst2019fairness,overdorf2018,greene2019,dencik2018,hoffmann2019fairness,pena2019,bennett2019}. 

This debate has happened against the backdrop of a long history of critical reflection on the values embodied in technical artifacts, and the need to design with such values in mind. Scholars have spent decades working to draw attention to the normative commitments expressed in and through technology. This call for recognition of values has been a core project of science and technology studies~\cite{winner1980,agre1997, jasanoff2006}, Values-in-Design~\cite{introna2000,friedman2019}, legal scholarship~\cite{reidenberg1997,lessig2009, citron2008}, and allied disciplines for years, and remains an active area of research. Algorithms have become objects of particular scrutiny over the past decade~\cite{gillespie2016}.

Within computer science, the machine learning and mechanism design communities have been particularly active in taking up these concerns. Both fields have been heeding the call for attention to values, politics, and ``social good'' more generally, holding more than twenty technical workshops and conferences between them on some variation of fairness, bias, discrimination, accountability, and transparency in the last five years (e.g.,~\cite{fat*network,abebe2018mechanism}).

And yet, these recent efforts have been met with concern that computer science has failed to target the right point of intervention. In focusing on changes to decision-making and allocation, these endeavors risk obscuring the background conditions that sustain injustices. For instance,  a computational intervention that aims to equalize offers of college admission across demographic groups might function as a less ambitious substitute for the deeper and more challenging work of improving high school instruction in low-income neighborhoods. Similarly, an intervention at the selection phase in an employment context might mask the impact of a hostile workplace culture or other barriers to long-term employee success --- problems for which other (and perhaps non-technological) responses might be indicated~\cite{barocas2014putting}.

There is a long-standing tension between strategies that seek to intervene incrementally within the contours of an existing social and political system and those that seek more wholesale social and political reform. Existing work has made clear how computational approaches may contribute to the former style of intervention. Here, we ask whether, and to what extent, computing can contribute to the latter style as well. We pose this question while recognizing the critical scholarship we have described above, and we emphatically reject the idea that technological interventions can unilaterally ``solve'' injustice in society --- an  approach some critics condemn as ``solutionism''~\cite{morozov-save-everything}. Our goal is to cut a path between solutionist and critical perspectives by describing potential roles through which computing work can support, rather than supplant, other ways of understanding and addressing social problems.

Meaningful advancement toward social change is always the work of many hands. We
explore where and how technical approaches might be \textit{part} of the
solution, and how we might exploit their unique properties as a route to broader
reforms. In this way, we seek to address two distinct audiences: this paper is a
call to our colleagues in computer science to reflect on how we go about our
work in this area, as well as a response to our fellow critical scholars who
question computing's value in addressing social problems. In what follows, we
propose a series of four potential roles for computing research that may be
well-aligned with efforts for broader social change.\footnote{These proposals
are in the spirit of other scholars' recent calls to recognize the political
nature of computational work and articulate new modes of engagement for
computing (e.g.,~\cite{green2019data}).}

The roles we offer here are intentionally modest: they are ways to leverage the particular attributes of computational work without overclaiming its capacity. We argue that these are worthy and plausible aspirations for our nascent field. Our list is illustrative, not exhaustive, and our categories are not mutually exclusive. And like any attempt to effectuate change, the approaches we outline carry hazards of their own, which we also explore below.\footnote{While we include some examples from low- and middle-income nations, many of the examples and hence discussions are focused around the United States and other developed nations. There is an emerging set of discussions around risks and challenges in using machine learning specifically in the context of developing nations, including the annual Machine Learning for Development Workshop Series~\cite{de2018machine}, which also faces many of the structural challenges we engage here.}

\section{Computing as Diagnostic}
\noindent \emph{Computing can help us measure social problems and diagnose how they manifest in technical systems.}

While computing cannot solve social problems on its own, its methods can be used to diagnose and precisely characterize those problems. Computational approaches, used in tandem with other empirical methods, can provide crucial evidentiary support for work that attends to values in technology --- even if computing itself is an insufficient remedy. In this sense, there is a role for computer science and its methods to play in critically interrogating computing itself. 

Many now-classic studies in the field take this approach, giving us a new sense of the shape and depth of a long-standing problem by applying a computational lens. Latanya Sweeney's analysis of an ad delivery platform demonstrated, among other things, that arrest-related ads were more likely to appear in response to searches for first names commonly associated with African-Americans~\cite{sweeney2013discrimination}, likely reflecting Internet users' disproportionate propensity to click on ads suggesting that African-Americans, rather than people of other races, have criminal records. Teams of computer science researchers have demonstrated the gender biases inherent in word embeddings and the difficulties they create for machine translation and other language tasks~\cite{caliskan2017semantics,bolukbasi2016man}. More recently, Joy Buolamwini and Timnit Gebru demonstrated that several commercially available facial analysis systems perform significantly worse on women and individuals with darker skin tones~\cite{buolamwini2018gender}. These analyses have enhanced our ability to document discriminatory patterns in particular sociotechnical systems, showing the depth, scope, and pervasiveness of such problems and illuminating the mechanisms through which they occur. 

And yet, while these studies sometimes point toward improvements that could be made (e.g., to ensure that ads appear at comparable rates across groups, to use caution with language models trained on text corpora with ingrained human biases, or to train facial analysis systems on a more diverse dataset), they do not purport to offer computational \textit{solutions} to the problems they illustrate. From this perspective, recent attempts to develop formal definitions of fairness, while often presented as a solution to bias, can be understood as diagnostics first and foremost --- as new methods and metrics for evaluating normatively relevant properties of computational systems~\cite[Chapter~2]{fairmlbook}. 

Diagnostic work of this sort can be particularly valuable when confronting systems that are ``black-boxed'': when the criteria of decision-making (and the values that inhere therein) are obscured by complexity, trade secret, or lack of transparency, diagnostic studies can provide an important means of auditing technical processes~\cite{sandvig2014auditing}. And such an approach can help to highlight technological dimensions of a social problem in a way that would be difficult to interrogate otherwise. For example, Abebe et al. leverage search data to surface unmet health information needs across the African continent --- demonstrating when and how search engines display low-quality health information to users. A lack of high-authority websites relevant to users' search queries might mean that results highlight natural cures for HIV, rather than scientifically sound medical advice~\cite{abebe2019using}. 

Other recent work has sought to advance the diagnostic role that computing can play by providing a principled approach to interrogating the machine learning pipeline~\cite{gebru2018datasheets,mitchell2019model}. These diagnostic efforts do not present themselves as solutions, but rather as tools to rigorously document practices. Thus, when compared with other computing interventions that aim directly at incremental improvements, they are less vulnerable to becoming a substitute for broader change. These efforts are not intended to absolve practitioners of the responsibility to critically examine the system in question, but instead to aid in that investigation.

Without purporting to resolve underlying social problems, diagnostic studies that use computational techniques can nevertheless drive real-world impacts on the systems they investigate. For example, in response to Buolamwini and Gebru's findings~\cite{buolamwini2018gender}, Microsoft and IBM reported that they had improved the accuracy of their facial analysis technologies along gender and racial lines~\cite{roach2018microsoft,puri2018mitigating}. Continued research (e.g.,~\cite{raji2019actionable}) and advocacy following this work has led to policy-making and government action that questions, limits, or prohibits the use of facial recognition in a number of contexts.

Some important caveats are warranted. First, computing is not unique in its capacity to help us diagnose social problems, even those manifest in technical systems. Disciplines like science and technology studies (STS), sociology, and economics provide their own sets of tools to interrogate sociotechnical phenomena --- including tools that capture important dimensions poorly addressed by computational approaches. For example, descriptive ethnographic research is essential for understanding how social and organizational practices intersect with technical systems to produce certain outcomes (e.g.,~\cite{brayne2017big,christin2018counting}. Computing is only one of multiple approaches that should be brought to bear on the analysis --- even when it may appear that the system in question (e.g., facial recognition technology) is primarily technical. A holistic analysis of a sociotechnical system must draw from a wide range of disciplines in order to comprehensively identify the issues at stake~\cite{crawford2016there,selbst2019fairness}.

Second, our optimism as to the potential benefits of computing as diagnostic must be tempered by a healthy degree of skepticism. Once metrics and formulae are used to characterize the performance of a system, individuals and organizations often have incentives to optimize towards those metrics in ways that distort their behavior, and corrupt the meaning of the metric. This process puts any targeted metric at risk of becoming a less useful measure over time --- a phenomenon known as Goodhart's Law~\cite{hoskin1996awful} or Campbell's Law~\cite{campbell1979}. We can see this worry instantiated, for example, in the Equal Employment Opportunity Commission's 4/5ths rule. The rule applies to employee selection procedures, and holds that when the selection rate for one protected group is less than 80\% of that of another group, such a gap is strong evidence of proscribed disparate impact. Initially intended as a guideline to aid in diagnosis of potentially discriminatory practices, it has become a target in itself for practitioners who strictly enforce the 4/5ths rule while sometimes failing to consider other aspects of discrimination and bias~\cite{raghavan2020mitigating}. Thus, when using computational techniques as diagnostic tools, care must be taken to prevent diagnostics from implicitly becoming targets.

Diagnostic approaches may be stymied when technical systems have been insulated from outside scrutiny~\cite{pasquale2015black}. For example, while a number of studies have used computational methods to characterize proprietary algorithms used in news curation~\cite{lurie2019opening}, resume search~\cite{chen2018investigating}, and ad delivery~\cite{datta2015automated,ali2019discrimination}, their methodologies must work around the constraints imposed by the institutions that operate these systems. Moreover, some testing methods can be further impeded by legal restrictions on researchers' activities (e.g., the disputed use of the Computer Fraud and Abuse Act to attach criminal penalties to audit methods that involve violating a website's terms of service~\cite{patel2018testing}). In such cases, it can be infeasible to provide a fully comprehensive or conclusive diagnosis.

Finally, it can be tempting to view diagnosis itself as a goal: precisely stating the problem, we might hope, will naturally lead to a solution. Indeed, a good diagnosis can motivate the relevant parties to work towards remedying the concerns it surfaces. The above examples, and others that follow in subsequent sections, demonstrate that illuminating a problem through technical examination can lead to positive change. However, it is important not to confuse diagnosis with treatment: there are a number of domains in which we are well aware of the extent of the problem (e.g., mass incarceration or homelessness), yet do not have sufficient will or consensus to adequately address it. As Ruha Benjamin notes, the production of audit data about a social problem must be accompanied by narrative tools that help to drive change: ``[d]ata, in short, do not speak for themselves and don't always change hearts and minds or policy''~\cite{benjamin2019race}.

\section{Computing as Formalizer}
\noindent \emph{Computing requires explicit specification of inputs and goals, and can shape how social problems are understood.}

People and institutions charged with making important decisions often speak in general terms about what they are doing. A standard that says that social workers must act in the ``best interests of the child,'' for instance, or an employer's stated intent to hire the ``most qualified applicant,'' leave wide latitude for interpretation. This vagueness is a double-edged sword. At its best, such broad flexibility lets decision-makers consider factors that are specific to a particular case or situation, and that could not be stated in a rigid rule. Vagueness is also an important political tool: laws are often worded broadly because no majority of legislators could agree on a more specific plan. Yet an underspecified standard may effectively delegate important choices to people and situations that are less open to scrutiny and supervision, and where individual bias may play a larger role. The relative merits of holistic and flexible ``standards'' versus more rigid and consistent ``rules'' are a prominent topic of debate among legal scholars~\cite{kaplow1992rules}. 

Increasing the role of computing within high-stakes decisions often moves decision-making away from generalized standards and toward more explicitly specified, rule-based models~\cite{citron2008}. A computational model of a problem is a statement about how that problem should be understood. To build and use a model is to make, and to promote as useful, a particular lens for understanding what the problem is. The nature of computing is such that it requires explicit choices about inputs, objectives, constraints, and assumptions in a system. The job applicant who is ranked first by a statistical model or other algorithm will not simply be the one holistically deemed most qualified; she will be the one ranked first on the basis of discrete numbers and specific rules. Those rules will reflect a range of concrete judgments about how job performance should be defined, measured, and forecasted.

This quality of computing is often bemoaned: Algorithms are cold calculators that collapse nuance and execute their tasks in ignorance of the histories and contexts that precede and surround them~\cite{noble2018algorithms,selbst2019fairness,eubanks2018}. But this same quality may also be a source of political potential. Because they must be explicitly specified and precisely formalized, algorithms may help to lay bare the stakes of decision-making and may give people an opportunity to directly confront and contest the values these systems encode~\cite{kleinberg2016guide}. In the best case, the need to disambiguate general policy aims may provide a community with important touchpoints for democratic deliberation within policy processes.

A model may be rigid in the way that it formalizes the structure of a problem, but the act of formalization is, paradoxically, a creative one, and increasingly important to social and political outcomes. In many settings where automated decision-making tools are used, this creative process of designing a system is a potential opportunity for non-technical stakeholders, including responsible officials and members of the public, to hash out different ways in which a problem \textit{might} be understood. Questions in this vein include: what is the specific predicted outcome that serves as the basis for decision-making, and why was it chosen over competing alternatives~\cite{passi2019problem}? Which outcomes should be treated as indicia of ``good'' job performance~\cite{barocas2016big}? How much risk should be considered ``high'' risk, meriting special treatment or attention from a court or a social services agency~\cite{koepke2018danger}? The process of formalization requires that someone make these decisions --- and gives us the opportunity to explicitly consider how we would like them to be made.

The formalism required to construct a mathematical model can serve as an opportune site of contestation, a natural intermediate target for advocacy. Transparency and accountability, moreover, are themselves intermediate virtues: Advocates will most often pursue these ends not out of an abstract interest in procedural questions, but rather as a way to influence the substance of decisions that the eventual algorithm will reach. Calls for transparency, accountability, and stakeholder participation --- while crucial in their own right --- ultimately do not resolve matters of substantive policy debate. These procedural values concern \textit{how} rules are made, but leave unaddressed \textit{what} those rules should actually be. The process of formalization can bring analytic clarity to policy debates by forcing stakeholders to be more precise about their goals and objectives.

This observation may seem out of step with the widely recognized risk that machine learned models may be so complex as to defy understanding. It may be difficult to succinctly describe how a model weighs different features or to provide a meaningful explanation for such a weighing. But all models are surrounded by technical and policy choices that are amenable to broad understanding and debate. The choice of objective function, for instance, and debates over which features to use as candidates for inclusion in the eventual model, are frequently delicate and politically important questions.

For example, using statistical models to assess people who have been accused of crimes is a controversial and widespread practice. These tools are described as ``risk assessment'' tools, but the probabilities that they actually forecast --- such as the chance that the accused person will miss a future appointment related to the case --- do not directly correspond to the narrow sense of ``risk'' that is legally salient at a bail hearing, namely the risk that the defendant will abscond from the jurisdiction or will violently harm another person before a trial can be held~\cite{koepke2018danger}. In situations like this, where the relevant risk is substantially narrower and more serious than the risks actually being measured and predicted by an algorithm, the algorithm effectively paints with too broad a brush, stigmatizing the accused by implicitly exaggerating the hazard that they pose~\cite{koepke2018danger}. Many failures to appear at a courtroom hearing are traceable to anodyne causes like a lack of transportation, or confusion about the time and place of the appointment. Formalizing the problem in a way that conflates these risks --- as some pretrial instruments do --- is a substantive and consequential decision, one that here operates to the detriment of the accused. For this among many other reasons, a broad coalition of more than a hundred civil rights organizations, as well as a substantial and growing number of technical experts on algorithms and criminal law, either oppose the use of these instruments in the forms they currently take or else have raised serious concerns about their use~\cite{riskassessmentstatement}.

The debate over how to formalize pretrial risk --- and over how and indeed whether to use these instruments --- is highly active at the time of this writing. Some advocates and scholars have argued that any feasible formalization of the problem, given the available data, will have drawbacks so serious as to draw into question the use of actuarial methods at all~\cite{mayson2019, 2018layers, riskassessmentstatement, koepke2018danger}. For instance, differing base rates of arrest across racial groups may mean that any feasible predictive system will amplify longstanding disparities by over-predicting arrests among stigmatized minority groups. Other activists, even while objecting to data-driven methods for these and other reasons, are also seeking to generate new kinds of data for the pretrial system to use, such as formalized and measurable evidence of the community support available to an accused person \cite{debug}. The inherent challenges of formally modelling this problem are driving \emph{both} changes to the models being used, and a reconsideration of whether risk assessment should be used at all.

In another context, Abebe et al.'s work~\cite{abebe2020subsidy} on income shocks demonstrates that alternative objective functions can result in very different allocations, even when the broad policy goal is the same. When allocating subsidies to households to prevent them from experiencing adverse financial outcomes due to income shocks, we might reasonably want to minimize the expected number of people that will experience a certain negative financial state (a min-sum objective function). Alternatively, we might instead reasonably determine that we want to minimize the likelihood that the worst-off family experiences this financial state (a min-max objective function). Abebe et al. show that these alternative policy goals --- both wholly plausible and defensible --- can result in some instances in the \textit{exact reverse} ordering of which households to prioritize for subsidies. Formalization, then, requires us to think precisely about what we really want when we say we want to accomplish some goal.

Of course, just because decisions must be made about how to formalize a problem does not mean those decisions will necessarily be made in a transparent, accountable, or inclusive way. There are many reasons why they may not be.

Some values may be harder to incorporate into formal models than others, perhaps because they are difficult to quantify~\cite{friedman1996bias}. What we choose to incorporate into models is largely driven by available data, and this constraint may press both scholars and practitioners to rely on measures that elide or distort important aspects of a situation, or that are not conducive to advocating for broader change. Although moral considerations are often important in determining system design, practical factors --- such as the cost savings involved in using readily-available data, rather than gathering or generating different evidence --- often play a vital role as well~\cite{clarke1988digital,citron2008,passi2019problem}. 

As a result, for stakeholders primarily committed to values that may be ill-served by the process of formalization, the decision to involve computing in a decision process may itself be a morally controversial one. The opportunity that formalization poses for constructive intervention has a negative corollary --- that pragmatic technical concerns can draw attention \textit{away} from the underlying policy goal that a model is supposed to serve. At the same time, researchers may be able to help illuminate the constraints and biases in existing sources of data --- such as child welfare ``substantiation'' outcomes, official findings of disability, or police arrest patterns --- and to underline both the importance and the expense of generating alternative data that may be a necessary precondition for socially responsible forecasting (e.g.,~\cite{wald1990risk}). 

If we fail to pay attention to the question of whether our formulation carefully reflects the true goals of the domain, models can easily fall prey to inaccurate assumptions or failures of imagination. In particular, the formalisms that a powerful institution adopts to judge individual people risk focusing moral evaluation and attention on individuals rather than on systems, a trend that some experts have long recognized as invidious. As far back as 1979, Donald Campbell argued that those who study public programs ``should refuse to use our skills in ad hominem research~\dots We should be evaluating not students or welfare recipients but alternative policies for dealing with their problems''~\cite{campbell1979}. More recent work has echoed this call for an evaluative focus on systemic ``interventions over [individual] predictions''~\cite{barabas2017}.

For instance, a discussion about the quality of employees in an organization might implicitly assume that the relevant metrics for each employee could be computed as a function of their attributes in isolation. This approach, however, would be unable to assess the quality of the team of employees that results, or the differing ways in which alternative candidates might alter the team's performance, since a focus on individual attributes would fail to model anything about interactions or complementarities among multiple employees~\cite{page-difference-book}.

\section{Computing as Rebuttal}
\label{sec:rebuttal}
\noindent \emph{Computing can clarify the limits of technical interventions, and of policies premised on them.}

In a world where practitioners are tempted to pursue computational interventions, technical experts have a unique contribution to make in illustrating the inherent limits of those approaches. Critical reflections on the limits of computing may drive some stakeholders --- and at times, the political process as a whole --- to reject computational approaches in favor of broader change. Technical experts can be particularly well positioned to recognize and declare these limits.
 
The belief that the design and deployment of computational tools is a neutral process --- and one that can be assumed to be beneficial --- may be long debunked in academic circles, but it remains a powerful force among policymakers and the general public. When scholars of computing recognize and acknowledge the political valence of their technical work, they can make it more difficult for others to leverage computing --- both practically and rhetorically --- for political ends. Technical experts can be particularly effective in contesting claims about technologies' capabilities and neutrality. This opportunity is the converse of the risk that the technical research community may ``fail to recognize that the best solution to a problem may not involve technology''~\cite{selbst2019fairness}, and that exploration of technological approaches may distract from broader aims.

For example: a group of computing scholars recently called on Immigration and Customs Enforcement (ICE) to reconsider its plans to use an algorithm to assess whether a visa applicant would become a ``positively contributing member of society'' as part of its ``Extreme Vetting'' program. The experts explained that ``no computational methods can provide reliable or objective assessments of the traits that ICE seeks to measure''~\cite{brennan2017letter}, and the program was later abandoned~\cite{harwell2018ice}. Other advocates --- drawing on computer science research --- have warned of the limits of natural language processing tools for analyzing social media posts for purposes of ``solving'' the problems of identifying fake news, removing extremist content, or predicting propensity for radicalization. The advocates noted that such tools are ``technically infeasible'' for such purposes and advised that ``policymakers must understand these limitations before endorsing or adopting automated content analysis tools''~\cite{duarte2018mixed}. Computing scholars and practitioners can play a critical role in advocating against inappropriate use of computational tools or misunderstandings of what they can accomplish.

Another mode of computing as rebuttal is work that uses a formalization of the underlying computational problem to establish mathematically rigorous limits on the power of algorithms to provide certain guarantees. One recent example comes from the study of prediction algorithms for risk assessment: formal analysis has shown that when two groups differ in their base rates for some behavior of interest, any well-calibrated algorithm assigning probabilities of this behavior to individuals must necessarily produce differences between groups in the rate at which people are inaccurately labeled ``high-risk'' or ``low-risk''~\cite{chouldechova2017fair,kleinberg2016inherent}. This result thus establishes that \textit{no matter what} algorithm is employed, any way of assigning risk estimates to two groups of differing base rates will necessarily produce some kind of disparity in outcomes between the groups; we cannot eliminate the problem through a better choice of algorithm --- or better process for making predictions, even those involving human judgment. Formal results of this type can expose the limitations of an entire category of approaches --- in this case, the assignment of risk scores to individuals --- in a way that the deficiencies of any one specific algorithm cannot.

Critiques of computing research can also illustrate the limitations of existing policy frameworks. Much computational research on fairness is built on frameworks borrowed from discrimination law --- for instance, the definition of protected categories, the 4/5ths rule as a metric for assessing disparate impact, and, perhaps most crucially, the belief that fairness can be achieved by simply altering how we assess people at discrete moments of decision-making (e.g., hiring, lending, etc.)~\cite{fairmlbook}. At best, discrimination law is an incomplete mechanism to remedy historic injustices and deeply entrenched structures of oppression. Work on computational fairness inherits these limitations~\cite{hoffmann2019fairness}. Yet, exposing the limits of algorithmic notions of fairness has exposed the limits of the underlying legal and philosophical notions of discrimination on which this work has built. In this regard, computer science has not simply taken normative concerns on board; it has helped to clarify and deepen our understanding of the values at stake~\cite{xiang2019legal,mayson2019,hellman2019measuring}.

Methods from computer science can also change how stakeholders perceive a problem domain, effectively serving as a rebuttal to entrenched policy thinking. For example, researchers who work on matching mechanisms for refugee resettlement have argued that by using more information about local communities --- for instance, matching refugees to places that already have a successful cluster of earlier immigrants speaking the same language --- they can improve resettlement outcomes and in turn increase the number of refugees that communities are willing to accept~\cite{kominers2018, kominers2016resettlement,jones2018matching}. Such steps are important because, as Robert Manduca has argued, ``many of our pressing social problems cannot be solved by better allocating our existing sets of outcomes''~\cite{manduca2019}. In some cases, computing can upend long-standing beliefs about the constraints under which policymakers are expected to allocate a scarce resource or opportunity. It can rebut needlessly narrow conceptualizations of a problem, expanding the set of outcomes considered achievable and the scope of the political debate. Conversely, by exposing the limits of what a more optimal allocation can accomplish on its own, computing can also help justify the \textit{non-computational} changes that are necessary for effective reform. A system for allocating scarce housing beds may make better use of a limited stock~\cite{eubanks2018}, and yet may ultimately highlight the need for more beds.\footnote{Eubanks~\cite{eubanks2018} discusses the January 2016 LA homelessness strategy report by Mayor Eric Garcetti which benefited from a previously-implemented coordinated entry system for allocating homelessness services in the city. This report was followed by policy changes that increased funding for low-income housing and homelessness services.}

Computing research has demonstrated the impossibility, infeasibility, or undesirability of proposed policies in other contexts as well. In the area of electronic voting, more than a decade of research and effort has demonstrated that secure, anonymous, and fully electronic voting systems are infeasible with present and reasonably foreseeable computer science methods. This view is now widely understood by policymakers, and has led large numbers of policymakers away from initial misplaced enthusiasm for paperless high-tech voting equipment. Similarly, an authoritative consensus among computing researchers about the infeasibility of secure ``backdoors'' in encryption technology has played a major role in dampening policymaker enthusiasm for broad surveillance mandates~\cite{abelson2015keys}.

Using computing as a tool for rebuttal carries risks. Perhaps the most significant is that proclamations about what a computational tool is \textit{incapable} of may focus attention on \textit{improving} said tool --- which may or may not be the desired policy outcome. This risk often accompanies critiques about inaccuracy and bias in predictive systems: if these critiques are perceived as the system not being (yet) ``good enough'' to accomplish designated policy aims well, technical improvements to the tool may be perceived as sufficient to address these problems. The danger, then, is that rebuttals run the risk of transforming a policy discussion into one about what is technically possible, rather than what social aims are ultimately desirable.

A decade ago, Kevin Haggerty observed this exact dynamic in debates about CCTV surveillance. The campaign against it had tried to push back against its adoption by pointing out its lack of efficacy. Haggerty noted that these debates were ``insidious precisely because they fixate exclusively on a standard of `functioning.' Such a narrow frame ignores the prospect that the true nightmare scenario might be that all of these technologies might actually work; that they might function perfectly, with full enrollment, complete transparency, seamless integration and exacting discriminatory power. Indeed, when critics point out that a surveillance technology does not work, one wonders if they would be thrilled if it did. Rather than confronting the dystopian potentials inherent in the maximum surveillance scenario, the political/methodological knife fight seems to unwittingly help drive systems towards developing ever-greater capacities''~\cite{haggerty2009methodology}.

Researchers focused on the risks of facial recognition seem to have avoided the same trap this round, raising alarm about \textit{both} its likely failures and threatening successes. Nabil Hassein was (to our knowledge) the first to publicly observe that working to equalize the accuracy of facial recognition across racial groups is an entirely different undertaking than seeking to restrict the use of surveillance based on facial recognition by law enforcement altogether~\cite{nabil2017}. The former approach addresses specific deficiencies in algorithms used by government actors and others, while the latter argues that the use of even a perfectly accurate facial recognition algorithm by police would be an amplifier for oppression and injustice. On this account, the best alternative to a bad algorithm might not be a better algorithm --- it might be no algorithm at all. Channeling these concerns, a number of scholars published an open letter calling on Amazon to stop selling its facial recognition technology to law enforcement altogether~\cite{concerned2019recent}. Several cities in the United States have issued bans against police use of facial recognition, with further bans under consideration across various states, at the federal level, and outside the United States.\footnote{For an updated list of jurisdictions that have already adopted or are considering regulations, see: \url{https://www.policingproject.org/face-rec-all-legislation}}

A second risk of the rebuttal approach is that focusing on what is \textit{not} possible may encourage policymakers to throw up their hands at a problem and unduly write off computational approaches altogether, when they may still have a positive role to play. Analysis of social media posts for hate speech, for example, is impossible to execute with complete fidelity; but it does not necessarily follow that platforms have no responsibility to try to filter such content. Rather, the incapabilities and imperfections of computational tools must be objectively and pragmatically considered when evaluating potential solutions. Scholars and practitioners engaging in computing as rebuttal must be cautious against overclaiming for this reason.

\section{Computing as Synecdoche}
\noindent \emph{Computing can foreground long-standing social problems in a new way.}

Virginia Eubanks's acclaimed ethnography \textit{Automating Inequality} tells the stories of three automated systems used to administer public services in different parts of the country~\cite{eubanks2018}. We learn about an attempt to automate welfare eligibility determinations in Indiana which wreaks havoc on the lives of the state's most vulnerable. In Los Angeles, the county's ``coordinated entry'' system ranks homeless individuals in order to apportion the city's limited housing supply --- treating basic human needs as resources to be strategically allocated, not rights to be ensured. And in Allegheny County, Pennsylvania, Eubanks tells the story of a risk model used to predict child abuse and neglect that targets poor families for far greater scrutiny.

Eubanks's book draws much-needed attention to these systems. But in discussing the book, and in the text itself, Eubanks is explicit that her core concern is \textit{poverty}, not technology. For Eubanks, computing is just one mechanism through which long-standing poverty policy is manifested. As she terms it, data analytics are ``more evolution than revolution'': they're a new instance of the tools we've used for generations, built on the same notions of desert and blame that have long characterized poverty discourse. Automated systems sit alongside broader social policies and cultural attitudes that divert poor people from the resources they need. Yet a technological lens does important work for Eubanks: by framing her book through algorithmic instantiations of poverty policy, she brings renewed attention to the plight of the poor writ large. It's safe to say that Eubanks's work has garnered more attention (and attention from different spheres of influence) as a book about inequality through technology than it might have if it were a book about inequality in general.

In this way, computing acts as a synecdoche --- a part that stands in for the larger whole in discourse and critique. Computing can offer us a tractable focus through which to notice anew, and bring renewed attention to, old problems. This approach is not uncommon in political discourse. Social problems are by nature complex and multivalent; we rarely obtain purchase on a problem by focusing on its entirety. In policymaking, we often chip away at big social issues by concentrating on their components --- and for both better and worse, technology critique often captures public attention. Even when we are ultimately concerned about a broader problem and see computing as but one (and perhaps not even the most important) facet of that problem, framing it as a technology problem may be quite pragmatic. Such a focus can leverage resources and attention that might not accrue otherwise. Put most bluntly, many people would not pick up a book about poverty policy in general --- but are game to read a critique of the algorithms used to administer it.\footnote{As Eubanks writes, the United States stands in ``cultural denial'' of poverty, ``a social process organized and supported by schooling, government, religion, media, and other institutions'' on both the left and the right~\cite{eubanks2018}.}

Jack Balkin describes this as a \textit{salience} function of new technologies. He proposes that, rather than focusing entirely on technological novelty, researchers should ask: ``What elements of the social world does a new technology make particularly salient that went relatively unnoticed before? What features of human activity or of the human condition does a technological change foreground, emphasize, or problematize?''~\cite{balkin2004digital}. In his own analysis of the Internet and free speech, Balkin emphasizes that digital technologies ``place[] freedom of speech in a new light'' and ``bring[] features of the system of free expression to the forefront of our concern, reminding us of things about freedom of expression that were always the case''~\cite{balkin2004digital}. In writing about Internet speech, Balkin tells us something about characteristics of speech more generally.

The significant risk, of course, is that a focus on the technological aspects of a problem can restrict our attention to merely those aspects. A computing lens can have the effect of masking and pulling political capital away from other and more insidious facets of a problem, as well as other (non-technical) means of addressing it.\footnote{Eubanks laments this narrowed horizon in the context of poverty policy: ``When the digital poorhouse was born, the nation was asking difficult questions: What is our obligation to each other in conditions of inequality? How do we reward caregiving? How do we face economic changes wrought by automation and computerization? The digital poorhouse reframed these big political dilemmas as mundane issues of efficiency and systems engineering: How do we best match need to resource? How do we eliminate fraud and divert the ineligible? How do we do the most with the least money?''~\cite{eubanks2018}.} It can also make computing something of a scapegoat; to reprehend computing for broad social wrongs may give us a convenient target for outrage at systemic injustice, but in a way that does not build momentum toward change. Further, using computing as a synecdoche strategically exploits computing's hegemony, but does not, at root, challenge it --- and may in fact reinforce tendencies to center the computational dimensions of problems while dismissing their other aspects.

Some of these tensions have recently emerged in critiques of the labor practices that underpin the operation of algorithmic systems. In Finland, prison laborers train a classifier on Finnish-language business articles; the startup running the project paid the prison the equivalent of what it would have paid Mechanical Turk workers for the same tasks. (It is not clear what proportion of the payment went to prisoners themselves.) Though the story was reported predominantly within tech media verticals, and the prison's practices decried by notable critics of technology, others noted that there was ``nothing special about AI''~\cite{chen2019inmates} in the story: its technological framing was merely a new illustration of long-standing practices. As Lilly Irani noted, ``[t]he hook is that we have this kind of hype circulating around AI'' as a gloss on ``really old forms of labor exploitation''~\cite{chen2019inmates}. Similar critiques focus on the working conditions of human content moderators who review violent, traumatic imagery for very low wages, and without proper support for the toll such work takes on their mental and physical health~\cite{roberts2019behind,gillespie2018custodians}. Low-wage and exploitative work existed long before computing; computing snaps them into focus and exacerbates them, but these are not merely computing problems, and we should not treat them as though they were.\footnote{In another labor context, concerns about automation in the workplace (of the ``robots will take all the jobs'' ilk) may also represent a technological scapegoat for deeper and more pernicious concerns about work precarity, capital accumulation, and the demise of affordable education and the social safety net. The latest waves of workplace automation may be exciting so much concern not because the technology is fundamentally different from prior advances, but because ``the programs that helped Americans deal with how technology is \textit{always} upending the job market were dismantled''~\cite{spross2019robots} (emphasis added). An overemphasis on robots obscures these broader problems and runs the risk of diverting further political capital from them.}

Sebastian Benthall makes a pointed critique on this theme. Channeling comments by Luke Stark, Benthall explains that, while debates about technology may at root be debates about capitalism, ``it's impolite to say this because it cuts down on the urgency that might drive political action. \dots The buzz of novelty is what gets people's attention''~\cite{benthall2018politics}. Rather than viewing a technology lens as a pragmatic route into a broader social or economic problem, Benthall contends that the AI ethics research community's focus on engineers and tech companies ultimately ``serves as a moral smokescreen'' that impedes effective critique of root problems. Benthall notes, for example, that the emphasis on AI ethics in the critical community is ``an awkward synecdoche for the rise of major corporations like Google, Apple, Amazon, Facebook, and Netflix''~\cite{benthall2018computational} --- and that the focus on technology rather than institutionalism or political economy necessarily circumscribes the issues that can be brought to the table.

To recognize that technology is only one component of a broader sociopolitical system is not to give technology a free pass. The existence of a technology may help to create the conditions to support and intensify a particular arrangement of power. Langdon Winner described this quality as technology's inherent compatibility with particular sociopolitical systems (noting, for instance, that nuclear power is inherently compatible with centralized and hierarchical decision-making as the social ``operating environment'' which allows the technology to function practically, whereas solar energy may be more compatible with distributed egalitarian systems of decision-making)~\cite{winner1980}. Content moderation systems seem naturally to invite exploitative labor practices based on the scale and speed they require and the nature of the content to be moderated; the technology does not just happen to be paired with these practices, but is a close fit for it. A synecdochal focus on computing must walk a pragmatic --- and tenuous --- line between overemphasis on technical aspects, on one hand, and due recognition of the work computing does to reinforce social systems, on the other.

\section{Conclusion}
Technical work on prediction and optimization in policy domains can sometimes seem to favor slow and incremental approaches to social change --- taking a side, without necessarily intending to, against broader and more sweeping redesigns of existing institutional arrangements. But as we have sought to show, technical work can also operate as a constructive ally for broad social change. In this paper, we have described four ways computing research can support and reinforce more fundamental changes in society and politics. Such research can help to diagnose problems, shape through formalization the ways people understand social problems and possibilities, illuminate the boundaries of what can and cannot be achieved by technical means, and make long-standing problems newly salient. And it can do so in ways that are compatible with taking less, rather than more, of today's social landscape for granted.

\bibliographystyle{ACM-Reference-Format} 
\bibliography{refs}

\end{document}